\documentclass[journal,comsoc]{IEEEtran}
\makeatletter\if@twocolumn\PassOptionsToPackage{switch}{lineno}\else\fi\makeatother

\usepackage{graphicx}
\usepackage[T1]{fontenc}
\usepackage{cite}

\ifCLASSINFOpdf
\else
\fi
\usepackage{amsmath}
\usepackage[cmintegrals]{newtxmath}
\usepackage{stfloats}
\usepackage{url,multirow,morefloats,floatflt,cancel,tfrupee}
\makeatletter

\AtBeginDocument{\@ifpackageloaded{textcomp}{}{\usepackage{textcomp}}}
\makeatother
\usepackage{colortbl}
\usepackage{xcolor}
\usepackage{pifont}
\usepackage[nointegrals]{wasysym}
\makeatletter

\def\mcWidth#1{\csname TY@F#1\endcsname+\tabcolsep}

\def\cAlignHack{\rightskip\@flushglue\leftskip\@flushglue\parindent\z@\parfillskip\z@skip}
\def\rAlignHack{\rightskip\z@skip\leftskip\@flushglue \parindent\z@\parfillskip\z@skip}

\@ifundefined{etal}{}{}

\usepackage{ifxetex}
\ifxetex\else\if@twocolumn\@ifpackageloaded{stfloats}{}{\usepackage{dblfloatfix}}\fi\fi

\AtBeginDocument{
\expandafter\ifx\csname eqalign\endcsname\relax
\def\eqalign#1{\null\vcenter{\def\\{\cr}\openup\jot\m@th
  \ialign{\strut$\displaystyle{##}$\hfil&$\displaystyle{{}##}$\hfil
      \crcr#1\crcr}}\,}
\fi
}

\AtBeginDocument{%
  \@ifpackageloaded{endfloat}%
   {\renewcommand\efloat@iwrite[1]{\immediate\expandafter\protected@write\csname efloat@post#1\endcsname{}}}{\newif\ifefloat@tables}%
}%

\def\BreakURLText#1{\@tfor\brk@tempa:=#1\do{\brk@tempa\hskip0pt}}
\let\lt=<
\let\gt=>
\def\processVert{\ifmmode|\else\textbar\fi}

\@ifundefined{subparagraph}{
\def\subparagraph{\@startsection{paragraph}{5}{2\parindent}{0ex plus 0.1ex minus 0.1ex}%
{0ex}{\normalfont\small\itshape}}%
}{}

\newcommand\role[1]{\unskip}
\newcommand\aucollab[1]{\unskip}
  
\@ifundefined{tsGraphicsScaleX}{\gdef\tsGraphicsScaleX{1}}{}
\@ifundefined{tsGraphicsScaleY}{\gdef\tsGraphicsScaleY{.9}}{}
\def\checkGraphicsWidth{\ifdim\Gin@nat@width>\linewidth
	\tsGraphicsScaleX\linewidth\else\Gin@nat@width\fi}

\def\checkGraphicsHeight{\ifdim\Gin@nat@height>.9\textheight
	\tsGraphicsScaleY\textheight\else\Gin@nat@height\fi}

\def\fixFloatSize#1{}
\let\ts@includegraphics\includegraphics

\def\inlinegraphic[#1]#2{{\edef\@tempa{#1}\edef\baseline@shift{\ifx\@tempa\@empty0\else#1\fi}\edef\tempZ{\the\numexpr(\numexpr(\baseline@shift*\f@size/100))}\protect\raisebox{\tempZ pt}{\ts@includegraphics{#2}}}}

\AtBeginDocument{\def\includegraphics{\@ifnextchar[{\ts@includegraphics}{\ts@includegraphics[width=\checkGraphicsWidth,height=\checkGraphicsHeight,keepaspectratio]}}}

\DeclareMathAlphabet{\mathpzc}{OT1}{pzc}{m}{it}

\def\URL#1#2{\@ifundefined{href}{#2}{\href{#1}{#2}}}

\def\UrlOrds{\do\*\do\-\do\~\do\'\do\"\do\-}%
\g@addto@macro{\UrlBreaks}{\UrlOrds}

\edef\fntEncoding{\f@encoding}

\makeatother

\newif\ifmultipleabstract\multipleabstractfalse%
\usepackage{tabulary}
\makeatletter
\AtBeginDocument{\@ifpackageloaded{longtable}{%
\def\LT@makecaption#1#2#3{%
  \LT@mcol\LT@cols c{\hbox to\z@{\hss\parbox[t]\LTcapwidth{%
    \sbox\@tempboxa{#1{#2: } #3}%
    \ifdim\wd\@tempboxa>\hsize
      #1{#2: }\textsc{#3}%
    \else
      \hbox to\hsize{\hfil\box\@tempboxa\hfil}%
    \fi
    \endgraf\vskip\baselineskip}%
  \hss}}}
}{}}
\makeatother

   \makeatletter
  \def\fig@textbf{\textbf}
   \AtBeginDocument{\renewcommand\floatc@plain[2]{\setbox\@tempboxa\hbox{{\footnotesize#1.}\footnotesize\hskip.5em#2}%
    \ifdim\wd\@tempboxa>\hsize {\fig@textbf{\footnotesize#1.}}\footnotesize\hskip.5em#2\par
        \else\hbox to\hsize{\hfil\box\@tempboxa\hfil}\fi}}
    \makeatother
  
\usepackage{float}

\newcommand{\texttildeapprox}{{\fontfamily{pcr}\selectfont\texttildelow}}

\begin{document}

%

        \title{On the Use of Quantum Entanglement in Secure Communications: A Survey}
      
\author{K~Shannon,
        E~Towe, and 
        O~Tonguz\thanks{K~Shannon is with Carnegie Mellon University, Pittsburgh,PA 
        15213-3890, USA. He can be reached at kshannon@andrew.cmu.edu}\thanks{E~Towe is with Carnegie Mellon University, Pittsburgh, PA,
        15213-3890, USA. He can be reached at towe@cmu.edu}\thanks{O~Tonguz is with Carnegie Mellon University, Pittsburgh, PA
        15213-3890, USA. He can be reached at tonguz@ece.cmu.edu}}

\maketitle 

\begin{abstract}
Quantum computing and quantum communications are exciting new frontiers in computing and communications. Indeed, the massive investments made by the governments of the US, China, and EU in these new technologies are not a secret and are based on the expected potential of these technologies to revolutionize communications, computing, and security.

In addition to several field trials and hero experiments, a number of companies such as Google and IBM are actively working in these areas and some have already reported impressive demonstrations in the past few years. While there is some skepticism about whether quantum cryptography will eventually replace classical cryptography, the advent of quantum computing could necessitate the use of quantum cryptography as the ultimate frontier of secure communications. This is because, with the amazing speeds demonstrated with quantum computers, breaking cryptographic keys might no longer be a daunting task in the next decade or so. Hence, quantum cryptography as the ultimate frontier in secure communications might not be such a far-fetched idea.

It is well known that Heisenberg's Uncertainty Principle is essentially a "negative result" in Physics and Quantum Mechanics. It turns out that Heisenberg's Uncertainty Principle, one of the most interesting results in Quantum Mechanics, could be the theoretical basis and the main scientific principle behind the ultimate frontier in quantum cryptography or secure communications in conjunction with Quantum Entanglement. In this survey paper, we provide a simple and accessible tutorial survey for engineers and computer scientists on the basics of Quantum Communications, foundations of Quantum Entanglement (QE), the mechanisms that can be used to generate QE, metrics to measure QE, Requirements for Channel Capacity and Noise, etc. While the primary application of QE considered is in Quantum Key Distribution for secure communications, other applications using entanglement protocols and their relevant considerations are also described.

Index Terms - Quantum Entanglement, Quantum Cryptography, Quantum Key Distribution, Quantum Communications, Quantum Computing, Secure Communications, Security.
\end{abstract}
    
%
\IEEEpeerreviewmaketitle

\section{Introduction}
\bigskip
This paper introduces the basic properties and practical uses of quantum entanglement when used for secure communications and quantum key distribution (QKD).

Broadly speaking, entanglement is used to create correlation between physical systems or parties. This could enable a number of protocols that can exploit enhanced correlation, especially in security applications. For example, quantum key distribution (QKD) is able to leverage correlation between two systems to create shared keys between two parties. In addition, Heisenberg's uncertainty principle guarantees that these keys will remain secret from malicious eavesdroppers.

In this paper, we describe the current methods and techniques used for generating entanglement. It will be shown that there is a tradeoff between easily created forms of entanglement and those types that would be most desirable. Entanglement generation has reached very high rates with high fidelity in wavelength and time in recent experiments and field trials and is now capable of megabyte per second rates on multiplexed channels. However, these entangled photons are not easily convertible to arbitrary multi-photon entangled states. It is important to understand the different types of entanglement, what they are each capable of, and what can be achieved with the available resources.

This survey paper is intended for communications engineers and practitioners who want a basic intuition and understanding for what Quantum Entanglement (QE) is good for, when it should be used, and what considerations to take into account when using QE. Rules of thumb for entanglement metrics, different protocols, requirements on channel capacity and noise are given to provide an overview of what is required in different settings. While the primary application of QE is in QKD, other applications and their relevant considerations are also described.

This paper emphasizes connecting the current practices in quantum communications, particularly QKD, to general intuition and guidelines for metrics of entanglement and noise. Different settings require different types of entanglement and different protocols to make the most use of the available quantum capacity. These are the applications this paper will discuss.

The remainder of this paper is organized as follows: Section II lays out the basic components of quantum communication: quantum states, entangled states compared to separable states, the role of measurement and what can be measured, metrics for entanglement, and models for quantum channels. Section III covers some typical applications of this description with some simple, yet important, ways to communicate over a quantum channel. Section IV covers the practical implementation of the abstract channels and protocols of Sections II and III. Section V is a summary of the major QKD protocols as well as some other quantum communication protocols. Section VI summarizes the key ideas and the intuition that are fundamental to working with quantum communication systems; this Section also makes some observations about the pros and cons of the general techniques. Section VII covers our observations of what fields of research are currently promising in the field. Section VIII discusses related surveys in this field, and Section IX makes some concluding remarks about the paper and potential directions for future work.

\section{Basics of Quantum Communication}
\bigskip

\textbf{What is Entanglement?}

In the context of communications, entanglement is best viewed as correlation between nonlocal measurements. By itself, it does not allow one to send information but only to obtain mutual information. So, to understand what entanglement is, we must understand how an entangled system differs from one that is independent, or only classically correlated. Systems are independent if knowledge of one system doesn’t give useful information about the state of the other. In the discussion below we follow the simplified approach provided by F. Wilczek in his insigthful paper [1].

\bgroup
\fixFloatSize{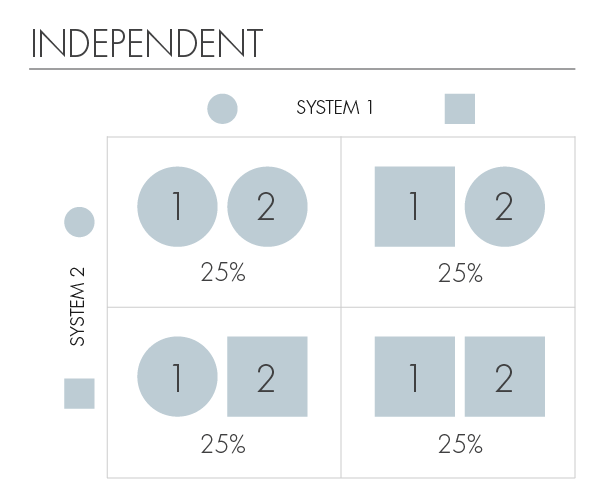}
\begin{figure}[!htbp]
\centering \makeatletter\IfFileExists{images/a8efacd2-cfc2-4e2a-b2ff-48ba54d534f8image1.png}{\includegraphics{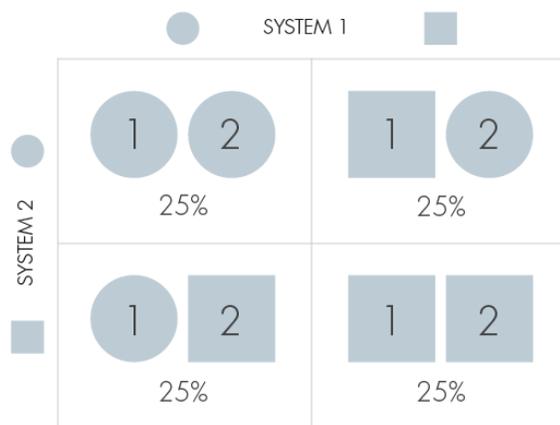}}{}
\makeatother 
\caption{{Independent, non-entangled systems\unskip~\protect\cite{Wilczek}.}}
\label{figure-c860c7f0a39740719fcfc52852e1217c}
\end{figure}
\egroup

In Figure 1, for example, measuring the shape of the first system (i.e., determining whether it is a circle or square) doesn’t give any information whatsoever about the second system and vice versa. As an example, if the first system were measured to be a square, the probability that the second system were a square would remain at 50\%. For entangled systems, things would be different as there will always be correlation of some kind. An example of this correlation is shown in Figure 2, where the shape measured in system 1 is always the same as system 2. In other words, if the first system were measured to be a square, the second would then be a square with 100\% probability. However, this alone is not sufficient to show entanglement, and will be the case for some classical systems as well. For example, if two lasers were set to have the same polarization, their photons would be correlated, but not entangled. While correlation is a necessary condition, the state of an entangled system also depends on what measurements are performed. Correlations in polarization, for instance, will not be specifiable beforehand, only once measurements are specified.

\bgroup
\fixFloatSize{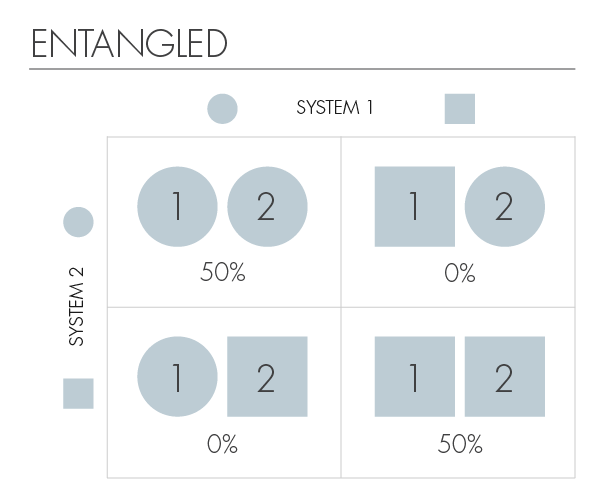}
\begin{figure}[!htbp]
\centering \makeatletter\IfFileExists{images/a8efacd2-cfc2-4e2a-b2ff-48ba54d534f8image2.png}{\includegraphics{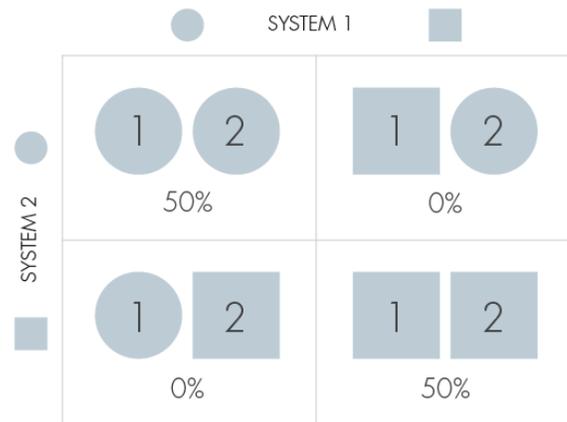}}{}
\makeatother 
\caption{{Entangled systems show correlation. However, correlation is not sufficient alone to show entanglement \unskip~\protect\cite{Wilczek}.}}
\label{figure-8f955815695e4ec784132c69fc989b20}
\end{figure}
\egroup

For a physical, two-party example, consider someone preparing two photons whose polarization is entangled, and then sending them in opposite directions down a fiber optic cable. Classically, when one photon is measured to be vertically or horizontally polarized, we know for certain that the other will be the opposite. However, the photons will not be opposite with respect to \ensuremath{\pm}45\ensuremath{^\circ} degrees polarization. A simple experiment can show that a photon with vertical polarization will randomly give \ensuremath{\pm}45\ensuremath{^\circ} when measured diagonally, so we know the photons can't be definitely vertical and +45\ensuremath{^\circ}, and so we have to conclude that they have no classical state, and that the state of one is affected by what measurement is performed on the other. This is shown in Figure 3, comparing the classical and entangled cases.

\bgroup
\fixFloatSize{images/Figure 3.png}
\begin{figure}[!htbp]
\centering \makeatletter\IfFileExists{images/Figure 3.png}{\includegraphics{images/Figure 3.png}}{}
\makeatother 
\caption{{Left: two photons are prepared with identical polarization, either vertical or horizontal. So the probability of measuring vertical or horizontal is 50\% for both parties. These measurements will be perfectly correlated. Alternatively, if both parties measure in the diagonal basis, the result will be \ensuremath{\pm}45\ensuremath{^\circ} with 50\% probability, but the results may not be identical.
\newline
Right: In the entanged case, the probability of each result is still 50\%, but the results will be identical regardless of what basis is chosen. Thus, the photons do not have a definite polarization until they are measured.}}
\label{figure-d725f2087e814c33ae87e194a5eba0cd}
\end{figure}
\egroup

The strange thing about this change in state is that it cannot be detected without knowledge of the full state. Heisenberg's uncertainty principle limits the amount of information that can be extracted from a quantum state. As measurements are made, the state is perturbed, destroying any remaining hidden information. This means that the measurements we choose to make are the key aspects of most quantum protocols; measurements cannot be performed without consequences and are the limiting factor, given a fixed amount of entanglement.

Measurement of an entangled state destroys some, or all, of the entanglement. Entanglement is best viewed as a physical resource, which is used up or "dissipated" as measurements are performed and exchanged for the type of communication we want to accomplish. Carefully choosing the type of measurements, making them quickly and precisely, and generating and preserving entanglement are the determining factors in how effective quantum entanglement based communication can be.

As a corollary to this fact, if and eavesdropper between Alice and Bob is performing measurements, they will destroy entanglement. Thus, Alice and Bob will not see the kind of correlation shown on the right side of Figure 3. Instead, the photons will look like simple "both up or both down" classical states, as shown on the left side. This makes entangled communication a way to detect interference with the signal. Hence, quantum entanglement based communications can be used for secure communications and secure QKD.

\textbf{Quantum Formalism}

Quantum communication is commonly described using qubits, the quantum analog of the classical 0 or 1 bit. A qubit can have states |0\ensuremath{\rangle} or |1\ensuremath{\rangle}, or superpositions of these two. For example, if we assign the vertical polarization of a photon to |0\ensuremath{\rangle} and the] horizontal polarization to |1\ensuremath{\rangle}, then a polarization of + 45\ensuremath{^\circ} will have a qubit value of |+\ensuremath{\rangle}= (|0\ensuremath{\rangle} + |1\ensuremath{\rangle}). When measuring in the horizontal/vertical basis, we are measuring in the |1\ensuremath{\rangle}/|0 \ensuremath{\rangle} basis of the qubit. A superposition of |+\ensuremath{\rangle} will return a result of |0\ensuremath{\rangle} or |1\ensuremath{\rangle} with equal probability.

If two photons are sent, we can send two qubits. The joint state can be |00\ensuremath{\rangle}, |01\ensuremath{\rangle}, |10\ensuremath{\rangle}, or |11\ensuremath{\rangle}, and the qubits can be measured separately. Joint states of systems like this can be written as the tensor product of the two states separately when using vector notation; however this can be written by appending the states in qubit form. For example, 

|00\ensuremath{\rangle} = |0\ensuremath{\rangle} \ensuremath{\otimes } |0\ensuremath{\rangle},

|0\ensuremath{\rangle} \ensuremath{\otimes } (|0\ensuremath{\rangle} + |1\ensuremath{\rangle}) = (|00\ensuremath{\rangle} + |01\ensuremath{\rangle}), and

|+\ensuremath{\rangle} \ensuremath{\otimes } |-\ensuremath{\rangle} = (|0\ensuremath{\rangle} + |1\ensuremath{\rangle}) \ensuremath{\otimes } (|0\ensuremath{\rangle} - |1\ensuremath{\rangle}) = 

(|00\ensuremath{\rangle} - |01\ensuremath{\rangle} + |10\ensuremath{\rangle} - |11\ensuremath{\rangle}) = |+-\ensuremath{\rangle}.

However, one can write out joint states that cannot be written as a tensor product of single qubit states. For example: | \ensuremath{\Psi } \ensuremath{\rangle} = | 00\ensuremath{\rangle} + |11\ensuremath{\rangle}, which could, for example, represent two photons whose polarizations are identical. This sort of state is called inseparable, and the two halves of a state are said to be entangled if and only if the joint state is inseparable.

In the classical example with hidden states of identical shape, the state could be written as |00\ensuremath{\rangle} or |11\ensuremath{\rangle}, with probability 50\% for each as shown in Figure 2. However, note that:

| \ensuremath{\Psi } \ensuremath{\rangle} = | ++ \ensuremath{\rangle} + | -- \ensuremath{\rangle}.

whereas |00\ensuremath{\rangle} = | ++ \ensuremath{\rangle} + | +- \ensuremath{\rangle} + | -+ \ensuremath{\rangle} + | -- \ensuremath{\rangle}.

So, there would be some chance of measuring diagonal polarization differently in the classical case, while the entangled case will always yield two identical results. Thus, the entangled system exhibits correlation beyond what is possible in a classical system.

Quantum measurement outcome probabilities are described by inner products, for example:

\ensuremath{\langle}0|0\ensuremath{\rangle} = 1, \ensuremath{\langle}0|1\ensuremath{\rangle} = 0

So if we measure the state |0\ensuremath{\rangle} in the |0\ensuremath{\rangle}/|1\ensuremath{\rangle} basis, the result is 0 with probability 100\%.

Quantum states can evolve locally in limited ways. The evolution of a quantum state is described by unitary operations. The key fact here is that such operations preserve inner products. Any inner product preserving operation is allowable, meaning measurements in the transformed basis is the same as measurements before the system evolves. In other words, if two states undergo the same transformation, they will maintain the same correlation.

A subset of unitary operations is called local quantum operations, and enable us to turn some states into others, like:

U: |0\ensuremath{\rangle} \ensuremath{\rightarrow}~|1\ensuremath{\rangle}, |1\ensuremath{\rangle} \ensuremath{\rightarrow}~|0\ensuremath{\rangle}

Then, applying U to the first photon of an entangled pair, the state is transformed:

U: |00\ensuremath{\rangle} + |11\ensuremath{\rangle}\ensuremath{\rightarrow} |10\ensuremath{\rangle} +~|01\ensuremath{\rangle}

In this way, we would call (|00\ensuremath{\rangle} + |11\ensuremath{\rangle}) and (|10\ensuremath{\rangle} +~|01\ensuremath{\rangle}) equivalent under local operations.

Note that some unitary operations are nonlocal, and allow for the generation of entanglement:

U: |00\ensuremath{\rangle} \ensuremath{\rightarrow}~|01\ensuremath{\rangle}, |01\ensuremath{\rangle} \ensuremath{\rightarrow}~|10\ensuremath{\rangle}, |10\ensuremath{\rangle} \ensuremath{\rightarrow}~|11\ensuremath{\rangle}, |11\ensuremath{\rangle} \ensuremath{\rightarrow}~|00\ensuremath{\rangle}

U: |00\ensuremath{\rangle} + |01\ensuremath{\rangle}\ensuremath{\rightarrow} |01\ensuremath{\rangle} +~|10\ensuremath{\rangle}

taking a separable state to a non-separable one.

An important operator will be the Hadamard transform (or H-gate), which takes |0\ensuremath{\rangle} and |1\ensuremath{\rangle} states to |+\ensuremath{\rangle} and |-\ensuremath{\rangle} states respectively. The Conditional NOT (CNOT) gate will flip the second qubit if and only if the first qubit is |1\ensuremath{\rangle}. Since |+\ensuremath{\rangle} and |-\ensuremath{\rangle} are superpositions, an H-gate followed by a CNOT gate will create entanglement.

\textbf{Types of Entangled Systems}

There are many kinds of entangled quantum systems, with different types and degrees of correlation. The simplest case is the Einstein-Podolsky-Rosen (EPR) pair: |01\ensuremath{\rangle} + |10\ensuremath{\rangle} where the state of each half is the opposite of the other. Practical examples of this kind of state are the spin of two electrons with an average of zero spin, or the momentum of a photon pair produced by Spontaneous Parametric Down Conversion (SPDC).

However, more than two electrons or photons can also be entangled. For example, in 3-party systems, one can have entanglement in the Greenberger-Horne-Zeilinger ( GHZ) state\unskip~\cite{703868:16682630}: |GHZ\ensuremath{\rangle} = |000\ensuremath{\rangle} + |111\ensuremath{\rangle}, or in the W state: |W\ensuremath{\rangle} = |001\ensuremath{\rangle} + |010\ensuremath{\rangle} + |100\ensuremath{\rangle} (named after W. D{\"{u}}r\unskip~\cite{703868:16682634}).The three -party entanglements have some fundamental differences from two-party states. For example, the correlations created by W and GHZ states will be very different; the W state is still entangled if any one of the three qubits is lost. However, the GHZ state becomes separable if any qubit is lost. So, these states each have their own application. For example, a GHZ state will always reveal the interference from an eavesdropper. A W state will not reveal how much information has been leaked, but by the same redundancy will be more robust to noise.

As seen before, a pair of entangled states can be transformed into another using only local operations. In general, an EPR pair can be transformed into any entangled pair of photons with local operations. However, W and GHZ states cannot be transformed into one another using local operations. They each belong to distinct classes of entangled three-party states. This is a new property of three-party states; any entangled two-party state can be created from EPR pairs and local quantum operations. The GHZ and W states together can create all other three - party states and form the canonical basis of three qubit entangled systems.

The difference between a GHZ and W state has practical ramifications: the GHZ state generates stronger correlations and is in this sense "more entangled." However, the W state is more robust to loss, and has a higher entanglement persistence. These notions will be made more precise in section IV, but the consequence is that there are several metrics for entanglement, the choice of which depends on the setting and the application at hand.

Later in the paper, we will discuss some protocols that go beyond the discrete |0\ensuremath{\rangle}/|1\ensuremath{\rangle} representation of quantum states, to consider measurements which can take on continuous values. For example, the length of time an entangled photon takes to be detected can be entangled with the measured energy of a second photon. These time/energy measurements will be continuous values with correlated Gaussian distributions. This correlation may be more sensitive to noise but can generate far more than one bit of shared information.

\textbf{Bell's Inequality and Entanglement Verification}

To see the correlation generated by entanglement, one can consider Bell's inequality. This inequality is a simple bound on the correlation of a classical system with definite properties and independent measurements. It states that if Alice measures properties Q and R, and Bob measures S and T, as shown in Figure 4, then:
\begin{center}
$ \textbf{E}[QS + RS + RT - QT] = \sum_{qrst}P(q,r,s,t)(qs+rs+rt-qt) \newline \leq \sum_{qrst}P(q,r,s,t)\times 2 = 2$
\end{center}

where E[] is the expected value of a measurement. In words, if Q and R both correlate with S, they must correlate with each other to some degree. For example, correlations of 85\% require an overlap of at least 70\%, or 2*p - 1. Similarly, if Q and R correlate with T, they also correlate with each other. The correlation of Q and R with S plus the correlation of - Q and R with T cannot sum to more than 1. However, the discussion up to this point has only dealt with a classical system. Entangled systems, as we will see, can violate this inequality due to the kind of correlations shown in Figure 3. Thus, Bell's Inequality will be used to prove entanglement.

Bell's Inequality assumes \textit{independence of the measurements} Alice and Bob perform. When this is not the case, the first equality shown above does not hold, as such a probability function as \ensuremath{P(q,r,s,t)} does not exist. The true probability function also depends on the measurements being performed: \ensuremath{P(q,r,s,t|M)}. A physical example is shown in Figure 4, for photon polarization.

\bgroup
\fixFloatSize{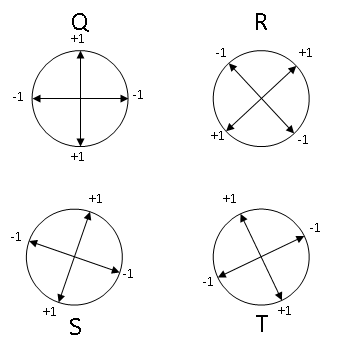}
\begin{figure}[!htbp]
\centering \makeatletter\IfFileExists{images/a8efacd2-cfc2-4e2a-b2ff-48ba54d534f8image5.png}{\includegraphics{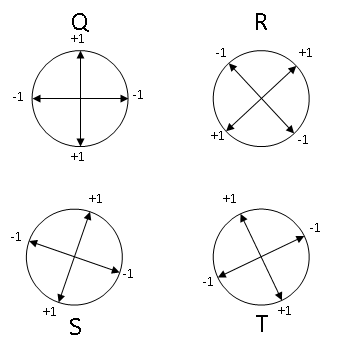}}{}
\makeatother 
\caption{{Suppose the polarization of two photons is entangled, such that their polarization is always identical (as shown in Figure 3). Now suppose Bob measures with basis Q or basis R, and Alice measures with basis S or T. They repeat the experiment many times and compare the average value of QS + RS + RT - QT to the Bell Inequality. \newline
As the polarizations are identical, measurements by Bob in the Q basis will have {\texttildeapprox}70\% correlations with measurements by Alice in the S basis. \ensuremath{So, E[QS] = .85}. The same will be true for RS, RT, and -QT. So, \ensuremath{E[QS + QT + RS - RT] = 2.4}, violating Bell's inequality, thus verifying quantum entanglement.}}
\label{figure-a4621fececaf4218b332824c6dd16111}
\end{figure}
\egroup

Bell's inequality gives one common way to verify entanglement. By measuring the observables Q, R, S, and T, one can check that the pair exhibits correlation beyond what is possible with a classical system. This can be used to ensure the security of protocols, as entanglement cannot be maintained by eavesdroppers. It also means that even if the entangled pairs come from an untrusted source, Alice and Bob can verify they are indeed entangled.

If entangled photons sent between two parties are being measured or entangled with by a third party in between, this can be detected. This is a consequence of the principle of "monogamy of entanglement," that if two systems are maximally entangled there can be no additional entangled states. Here, maximally will often practically mean how greatly the Bell Inequality is violated. The EPR and GHZ states are maximally entangled, as measured by their entropy. Thus, they will be easily found to violate the Bell Inequality or similar limits. However, the W state is not, and thus could be partially measured while maintaining entanglement. This illustrates the tradeoff between the robustness offered by the W state and the loss of security in using it; these are both measures of the amount of environmental correlation this state can take on without losing entanglement coherence. In general, the more persistence an entangled state has, the less sure one can be of whether it has interfered with the environment.

Rule of Thumb: There is a tradeoff between entanglement persistence and information about environmental interaction\unskip~\cite{703868:16682657}. One can formulate this as: 

"if I am not affected by environmental interference, I don't know about it."

\textbf{Entanglement Metrics}

One may care about many properties of a quantum channel. For example, we may care that at least some entanglement is preserved so that we can communicate in simple ways. Alternatively, it might be important to successfully transmit some large quantum state, like the GHZ state, in which case a higher probability of complete success is desirable. There are many ways to measure the entanglement of quantum states, which quickly diverge as more parties are added to the system. For 2-party states, the common and most useful metric relates all states to the EPR pair. This is because all states can be made from an EPR pair using local quantum operations, and EPR pairs can be distilled from other states.

The former process corresponds to entanglement of formation, defined as the ratio of EPR pairs to the number of target states out. For example, if 7 EPR pairs could make 10 copies of some state, that state would have an entanglement of formation of 0.7.

The latter process is known as distillable entanglement, defined as the ratio of EPR pairs which can be created to number of state copies put in. For example, if 7 EPR pairs could be made from 10 copies of the same state, that state would have a distillable entanglement of 0.7.

For 2-party states, the entanglement of formation is exactly equal to the distillable entanglement. Two-party states will frequently exhibit this kind of simplicity, as the EPR pair is the only canonical state from which all others can be made. Most measures of entanglement, information, or entropy will coincide for 2-party systems, but it is not so for 3 or more parties.

The entropy of a classical system is given by the well-known Shannon entropy equation \unskip~\cite{shannon}:
\begin{center}
$S_1(p) = -\sum^N_{\mu = 1} p_\mu \log p_\mu$
\end{center}

Where p denotes the full vector of probabilities, and each p\textbf{\ensuremath{_{\ \ensuremath{\mu\ }}}} denotes the probability of a given outcome. The higher dimensional counterpart of Shannon entropy is the R{\'e}nyi entropy\unskip~\cite{703868:16682657}:
\begin{center}
$S_q(\vec{p}) = \frac{1}{1-q}\log\left ( \sum^N_{\mu = 1} p_\mu^q \right )$
\end{center}

Where p and p\textbf{\ensuremath{_{\ \ensuremath{\mu\ }}}} are the same as above, and q is the order. Shannon entropy is given as q approaches 1, and the usual R{\'e}nyi entropy is given by q = 2. Note that Shannon entropy is E[log(p)], whereas R{\'e}nyi entropy is log(E[p]). When applied to quantum states, the R{\'e}nyi entropy (called the R{\'e}nyi{\textendash}Ingarden{\textendash}Urbanik entropy ) gives a measure of the level of entanglement. The lower the entropy, the higher the correlation of a state. One should consider the entropy over all unitary transformations, as states correlating in one basis may fail to correlate at all in another, revealing a lack of entanglement. Supposing that the state is non-separable, the lowest entropy over all unitary transformations would grant the highest entangled correlation

For example: in three dimensions the two states of canonical interest are the W and GHZ states. The R{\'e}nyi entropy of the W state is log( 3), whereas the GHZ state's is log(2). The smaller entropy of the GHZ state indicates a greater degree of entanglement.

\bgroup
\fixFloatSize{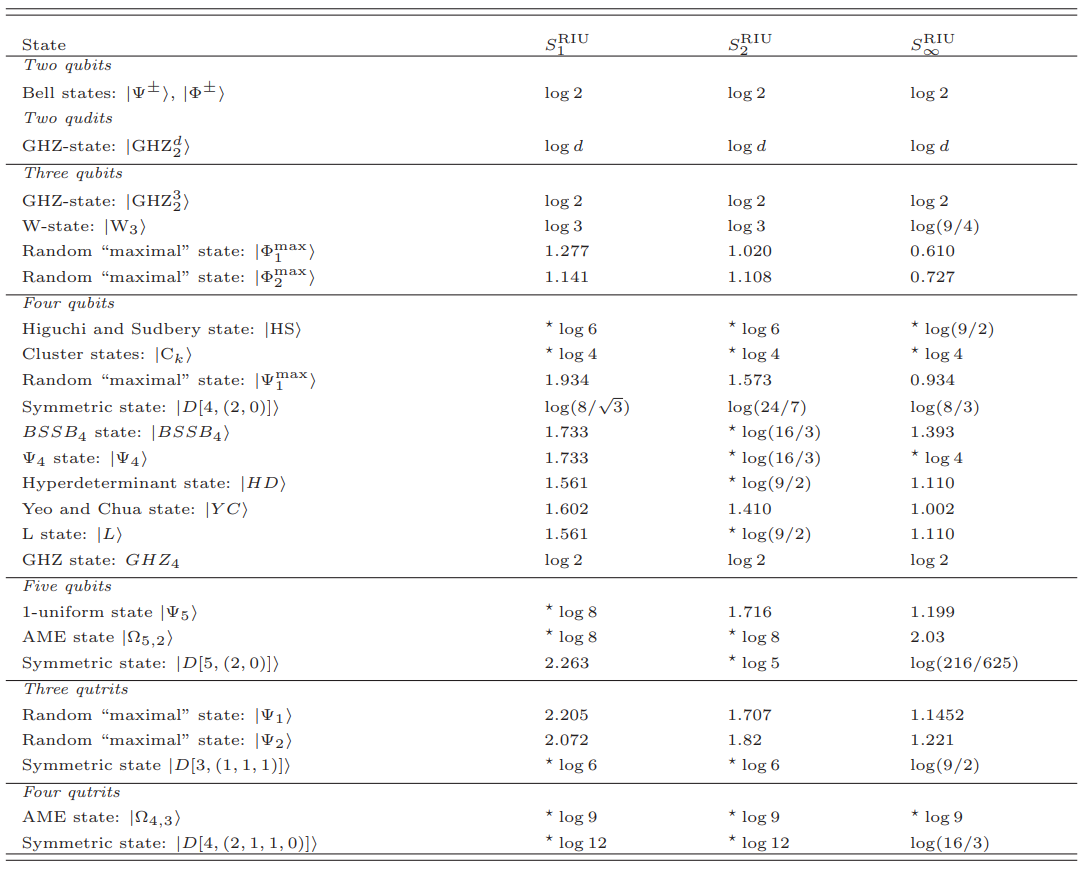}
\begin{table*}[!htbp]
\caption{{Entropies of order 1, 2, and infinity for different numbers of entangled parties\unskip~\protect\cite{703868:16682643}}}
\centering \makeatletter\IfFileExists{images/a8efacd2-cfc2-4e2a-b2ff-48ba54d534f8image8.png}{\includegraphics{images/a8efacd2-cfc2-4e2a-b2ff-48ba54d534f8image8.png}}{}
\makeatother 
\label{figure-682998ed516a405384872500b4da9099}
\end{table*}
\egroup
~For dimensions greater than 2, there is no analytical technique to minimize this equation over all unitary transformations, meaning characterizing high dimensional entanglement can be computationally intractable.

Another metric used is the Schmidt Measure of a state\unskip~\cite{703868:16682643}. If a state can be separated into a tensor product of R states:
\begin{center}
$C_{i_1, i_2, ..., i_K} = \sum^R_{\nu =1}\gamma_\nu a^\nu_{i_1} \otimes b^\nu_{i_2}\otimes ... k^\nu_{i_K}$
\end{center}

Then its Schmidt measure is log(R). This is the rank of the tensor for a state. The Schmidt measure gives an upper bound for R{\'e}nyi entropy, and may be easier to find in some cases.

An intuitively appealing measure is the ( Fubini-Study) distance to the nearest separable state\unskip~\cite{703868:16682643}. This is given by:
\begin{center}
$S^{RIU}_\infty(|\psi\rangle) = -\log F_{max}
\newline
F_{\max}(\psi) = \max_{U_{loc}}|\langle\psi|U_{loc}|0\rangle^{\otimes K}|^2$
\end{center}

This is the logarithmic geometric measure of entanglement. If instead one uses 1 - F\ensuremath{_{\ max}}, this is the linear geometric measure of entanglement.

For example, to find the Schmidt measure of a two -qubit state:
\begin{center}
$|\psi \rangle = \frac{1}{\sqrt{2}}(a|00\rangle + b|01\rangle + c|10\rangle + d|11\rangle) \;\;\;\;\;\;\;\; C = \begin{pmatrix}
a & b\\ 
c & d
\end{pmatrix}$
\end{center}

we can use matrix decomposition to write the state as a sum of tensor products:
\begin{center}
$C = UDV^T
\newline
|\psi\rangle = \sum_{j, \alpha, \beta} \sqrt{\lambda_j}|\alpha\rangle_1 U_{\alpha j} \otimes | \beta \rangle_2 V_{\beta j}$
\end{center}

so that in the Schmidt measure, R is the number of non-zero eigenvalues. Note that for a separable state, the determinant of C is 0, and for fully entangled state the determinant is 1.

These measures involve optimizing over all unitary transformations, which will have to be done numerically for any dimension larger than 2. The size of this problem makes it undesirable, and simpler metrics have been devised which are practical and interpretable.

The persistence of an entangled state is defined as the minimum number of local measurements required to destroy all entanglement. Among 3 parties, the GHZ state has persistence 1, while the W state has persistence 2, despite the GHZ state having a lower R{\'e}nyi entropy.

Given the ease of measuring entropy in bipartite states, one might devise metrics reducing to measuring bipartite entanglements. By partitioning the parties of a state into two groups, one can find the entropy between the two halves easily and unambiguously. Averaging over all such bipartite entanglements gives the Mayer-Wallach measure\unskip~\cite{703868:16682643}:
\begin{center}
$E_x(|\psi_K\rangle) = \frac{1}{L_K}\sum_A S_x(\rho_A)$
\end{center}

Generally, these metrics will scale logarithmically with the number of qubits, however this will depend on what type of state is chosen at each level. The number of states to choose from will scale super-exponentially, as there are 2\ensuremath{^{ n}} possible strings of n binary digits, and so$2^{2^{n}} $possible entangled states.

\textbf{Channel Capacities}
A quantum channel is anything that transmits quantum information, in the same way a classical channel transmits classical information. A quantum channel may refer to a free space channel or a fiber-optic channel. However, it may also be used to describe how well quantum information is transmitted through time, or from one system to another.

The capacity of a quantum channel can be described by any of the measures discussed below, with examples shown in Table I. Depending on the kind of entangled states we wish to end up with, different capacities may be appropriate. Maximizing the data rate of one protocol will be very different from another. The input-output entropy of states going through a quantum channel describes how much information is carried through. However, this will not be an elegant notion of capacity; channels will be non-additive, and inconsistent across metrics. For example, two channels of zero capacity can be used together to create a channel of positive capacity. Thus, characterizing the quantum capacity of a channel is usually specific to the application, in terms of Quantum Bit Error Rate (QBER).

The QBER of a protocol is a generalization of the classical bit error rate of noisy classical channels. In simple protocols this is sufficient, particularly for discrete variable protocols sending 0s and 1s. However, proving that a certain QBER is secure, in the sense that one can bound the information shared with the environment, can be difficult for more sophisticated protocols, as QBER becomes less universal as a metric. In addition, it is desirable to use homodyne detection to receive continuous variables. However, these cannot be described by a bit error rate as their alphabet size will be far larger. 

The classical capacity of a channel is a measure of how many classical bits can be sent by a quantum channel. For separable states, simple results apply. The Holevo Bound\unskip~\cite{703868:16682657}:
\begin{center}
$I(A:B) \leq S(\rho_A) - \sum_i p_i S(\rho_i)\equiv \chi$
\end{center}

reflects the fact that only one classical bit can be sent per qubit. Here, A is the bit string being sent and B is the measurement made by the receiver, and S is the Shannon entropy. This is because any measurement changes the state of a quantum system, destroying any residual information. For a noisy channel, the Holevo-Schumacher-Westmoreland Capacity\unskip~\cite{703868:16682624} gives the classical
capacity if entanglement is not allowed:
\begin{center}
$C(\mathcal{N}) = \max_{all \: p_i, \rho_i} \chi = \max_{all \: p_i, \rho_i} \ \left [ S(\sigma_{out}) - \sum_i p_i S(\sigma_i) \right ] \newline 
= \max_{all \: p_i, \rho_i} \left [ S \left ( \mathcal{N} \left ( \sum_i p_i \rho_i \right ) \right ) - \sum_i p_i S(\mathcal{N}(\rho_i)) \right ]\newline
= \chi(\mathcal{N})$
\end{center}

with all variables defined as above, and \ensuremath{\sigma  } refers to the qubits being sent. However, when entanglement is allowed, the classical capacity of a channel will increase exponentially, as will be shown in the next section. While it may not seem necessary to send classical bits over a quantum channel, this will give us the ability to send these bits secretly and will increase our understanding of what quantum communication is capable of.

\section{Basic Techniques in Quantum Communication}
\bigskip

\textbf{Superdense Coding and Teleportation}

There are two fundamental techniques in using entanglement in quantum communication:

\ensuremath{\bullet}  superdense coding

\ensuremath{\bullet}  quantum teleportation.

\bgroup
\fixFloatSize{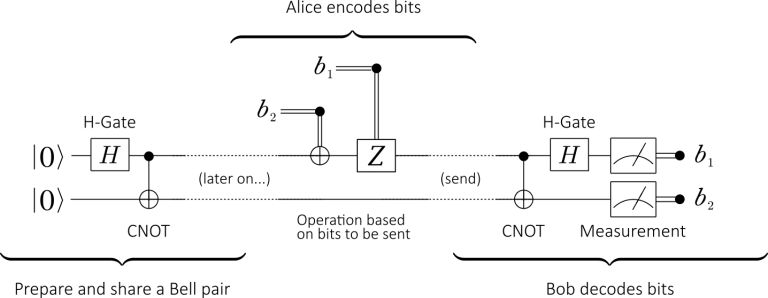}
\begin{figure*}[!htbp]
\centering \makeatletter\IfFileExists{images/a8efacd2-cfc2-4e2a-b2ff-48ba54d534f8image19.png}{\includegraphics{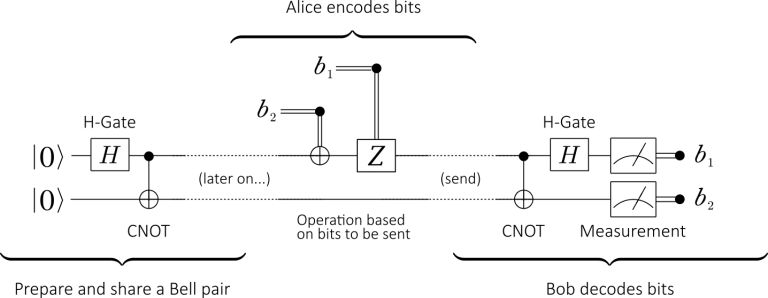}}{}
\makeatother 
\caption{{Gate diagram for superdense coding. The Hadamard gate (H-gate) takes |0\ensuremath{\rangle} states to |+\ensuremath{\rangle}, the Conditional NOT (CNOT) takes |00\ensuremath{\rangle} to |00\ensuremath{\rangle} and |10\ensuremath{\rangle} to |11\ensuremath{\rangle}. Thus, these gates create an entangled EPR pair, |00\ensuremath{\rangle} + |11\ensuremath{\rangle}. After Alice encodes the bits (\ensuremath{b_1} and \ensuremath{b_2}) by operating on the first qubit, Bob is able to perform a similar transformation and then measure the correct bits. Note that encoding happens on only one qubit, which must be sent, and two bits are received \unskip~\cite{quantum}.}}.
\label{figure-bfc71931cf4b44c69ccbd190dd214ed7}
\end{figure*}
\egroup

While the former uses entanglement to send classical information, the latter uses entanglement to send quantum information.

In superdense coding, an entangled state is shared between two parties, Alice and Bob. Alice performs local quantum operations on her half of the state, encoding the bits she wants to send. She then sends her half of the state to Bob through a quantum channel. Bob then measures the entire state, both his and Alice's half. The result he gets will indicate what bits Alice sent. This process is shown in Figure 5. Thus, for example, one qubit of communication can result in the transmission of two classical bits. In general, n qubits of quantum communication can yield 2\ensuremath{^{ n}} bits of information, if general entangled states are allowed. What is more, this method of communication is secure. If an eavesdropper intercepts the qubits Alice sends, she will not be able to perform the necessary measurement to decode the information. This illustrates the appeal of entanglement- based communication: even if noise restrictions are higher, a greater number of bits can be sent per qubit, balancing the loss.

In quantum teleportation, a previously distributed entangled state is again used, this time to send quantum information. Alice interacts with the qubits she wishes to send with her half of the state. She then measures the system which was originally entangled with Bob's, effectively switching the qubits to be sent with her half of the entangled state. She classically communicates her measurement result to Bob, who performs the corresponding quantum transformation to his half of the state. As a result, Bob has the desired qubit, and entanglement is lost. The process is shown in Figure 6. This technique can be used to enable quantum communication in a much further range than would otherwise be possible. For example, Alice could have an entangled pair with Bob, and Bob with Charlie. By teleporting his half of Alice's state to Charlie, Bob can allow Alice and Charlie to share an entangled pair. Thus, quantum communication of an unknown, unobserved quantum state can happen between two people who never send each other any quantum information. This is called entanglement swapping\unskip~\cite{703868:16682624}.

\bgroup
\fixFloatSize{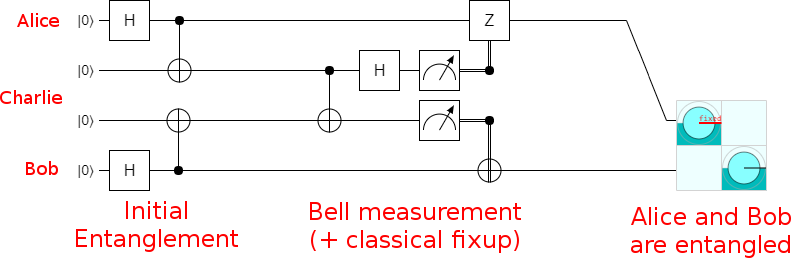}
\begin{figure}[!htbp]
\centering \makeatletter\IfFileExists{images/a8efacd2-cfc2-4e2a-b2ff-48ba54d534f8image20.png}{\includegraphics{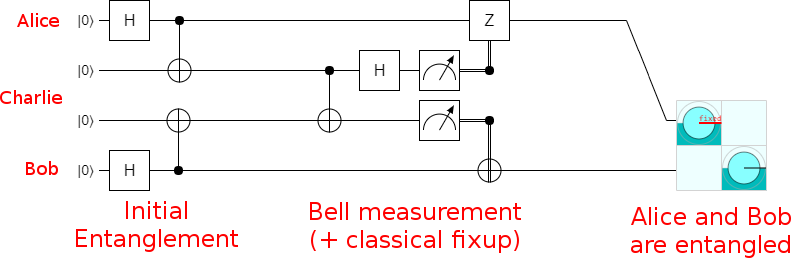}}{}
\makeatother 
\caption{{Similar to Figure 5, the Hadamard (H gate) and CNOT gates create entangled pairs, by the steps |00\ensuremath{\rangle} \ensuremath{\rightarrow}|00\ensuremath{\rangle} + |01\ensuremath{\rangle}\ensuremath{\rightarrow} |00\ensuremath{\rangle} + |11\ensuremath{\rangle}. Charlie performs a similar transform, and then by measuring his state, directs Alice and Bob how to correct their qubits. The result is that CHarlie is able to swap entanglements so that Alice and Bob share a pair, without having communicated \unskip~\cite{phys}.}}
\label{figure-66774592abe343d6814d79ff856052b5}
\end{figure}
\egroup

So, entanglement is able to turn quantum communications into (higher dimension, secure) classical communication, and can turn classical communication into quantum communication, provided that the proper transformations and measurements are possible. This flexibility of entanglement is a large part of its appeal, as a relatively simple infrastructure for distributing entangled states could allow a wide range of methods of communication.

\textbf{Basic QKD Schemes}

Quantum Key Distribution (QKD) is the process of using quantum signals to create a shared secret key between two or more parties. Such a process allows communicating parties to encrypt messages with said key and communicate securely from eavesdroppers. The advantage of using quantum signals, for example polarized photons, is that an eavesdropper on the key exchange can be detected. For example, if an eavesdropper measures the polarization of an entangled pair, then sends a new photon, entanglement will be destroyed and correlation won't hold in every basis. Thus, if QKD is successful, it is guaranteed that the key truly is secret and unbreakable.

\textbf{Review of} \textbf{ BB84 and E91 as basic QKD}

\bgroup
\fixFloatSize{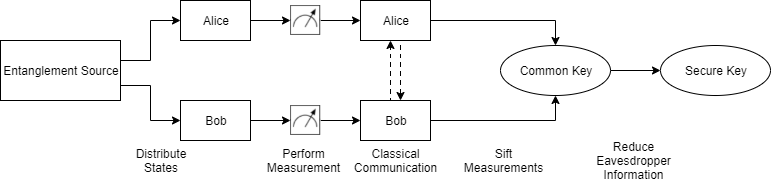}
\begin{figure*}[!htbp]
\centering \makeatletter\IfFileExists{images/a8efacd2-cfc2-4e2a-b2ff-48ba54d534f8image21.png}{\includegraphics{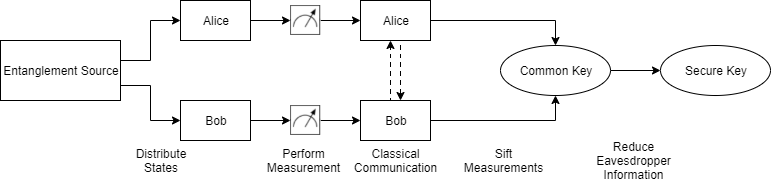}}{}
\makeatother 
\caption{{Block diagram of a general QKD process, where solid lines represent quantum channels and dotted lines represent classical channels. An entangled pair is distributed to Alice and Bob as in step ii. The quantum states could be of many types, and do not need to be entangled for protocols like BB84. Alice and Bob perform measurements (step iii), carry out the necessary classical communication of their bases and results (step iv), sift the key for errors (step v), then perform mixing and correction to prevent eavesdropping (steps vi and vii). This is a very abstract view of QKD, but these are the elements common to all of them.}}
\label{figure-8f11ddf195274087a7ec4cae10eec2bf}
\end{figure*}
\egroup

\begin{table*}[bp]
\caption{{Illustration of how BB84 works. Bases for BB84 are randomly chosen, matching choices result in matching bits. Unmatching choices are thrown away. The resulting key here would be 01001.}}
\label{table-wrap-c8ed78437540479caddc2d21cb4ffc0f}
\def\arraystretch{1}
\ignorespaces 
\centering
\begin{tabulary}{\linewidth}{LLLLLLLLL}
\hline 
 Bob's Basis &
   0/90 &
   45/135   &
   0/90 &
   0/90 &
   45/135   &
   0/90 &
   45/135   &
   45/135  \\
\hline 
 Alice's Basis &
   0/90 &
   0/90 &
   0/90 &
   45/135   &
   45/135   &
   0/90 &
   0/90 &
   45/135  \\
\hline 
 Bob's Encoding &
   0 &
   1 &
   1 &
   1 &
   0 &
   0 &
   1 &
   1\\
\hline 
 Measurement &
   0 &
   0 &
   1 &
   1 &
   0 &
   0 &
   0 &
   1\\
\hline 
 Result &
   0 &
   Discard &
   1 &
   Discard &
   0 &
   0 &
   Discard &
   1\\
\hline 
\end{tabulary}\par 
\end{table*}

All QKD schemes follow the same basic steps, illustrated in:

i. Everyone agrees on a protocol to run over an authenticated public classical channel.

ii. Some party prepares and sends quantum states to the others.

iii. All parties perform the necessary quantum measurements, obtaining some variables.

iv. Parties classically communicate the necessary information to align correlation, e.g., chosen basis, obtained observable, etc.

v. Using this information, they sift or otherwise change their variables to correct their keys to be exactly equal.

vi. Parties measure the amount of error and assess an eavesdropper's maximum mutual information.

vii. They then correct errors and amplify their privacy to the appropriate level.

The flow of quantum and classical communication in such a scheme is shown in Figure 7.

The first QKD protocol was proposed by Bennett and Brassard in 1984 and is known as BB84. In this protocol, Alice randomly encodes zeros and ones in a random choice of basis. Bob randomly chooses his basis to measure it and obtains a binary string. When Alice and Bob happen to choose the same basis, their bit will agree. Thus, after sending all the bits, Alice announces which bases she chose, and Bob confirms which ones were correct. The rest of the bits are uncorrelated, and so are thrown away (see Table II). However, knowing the basis used does not give an eavesdropper information about the bits, and so is not useful.

\bgroup
\fixFloatSize{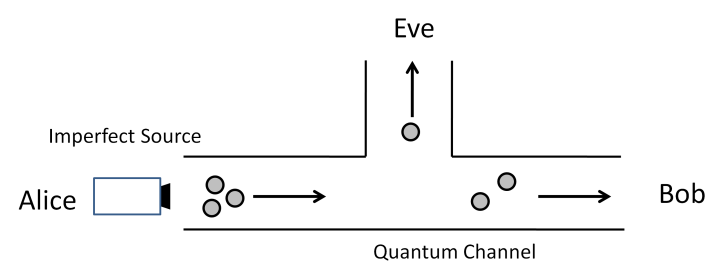}
\begin{figure}[!htbp]
\centering \makeatletter\IfFileExists{images/a8efacd2-cfc2-4e2a-b2ff-48ba54d534f8image22.png}{\includegraphics{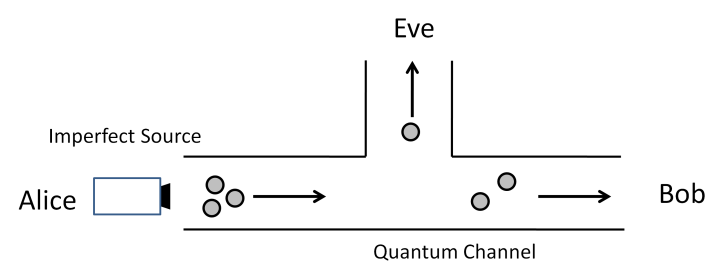}}{}
\makeatother 
\caption{{Eve can split signals containing multiple duplicate photons in order to gain information undetected}}
\label{figure-988de035674e4934aa560f1871b59573}
\end{figure}
\egroup

If someone, Eve, for instance, were to eavesdrop on this channel, she would have to guess the basis used. Since single photons are being sent, the signal cannot be partially measured; it must be either fully measured or not. Thus, by eavesdropping, Eve disturbs the polarization of the photons. When Alice and Bob's bases agree, but their bits do not, they conclude that someone must have disturbed the signal. They then try again on a different physical channel.

This protocol was modified in 1991 by Eckert, who proposed using Bell's inequality to verify that a channel was secure. It was subsequently shown by Bennett and Brassard that Bell's inequality was unnecessary, and that the exact same process as BB84 could be used by measuring one half of the pair and verifying on the other end as long as the bases were independent.

The main drawback of the original BB84 protocol is that it requires a quantum channel between Alice and Bob. In addition, it imposes significant practical challenges by requiring Alice to send only single photons and Bob to detect the single photons. Real light sources typically emit bunches of photons at a time, and single photon detectors are expensive to build and maintain at any reasonable detection efficiency.

The protocols using quantum entanglement no longer require Alice to send her own quantum signal, and entanglement verification like Bell's inequality allows them to verify security. However, these protocols have their own set of problems, most prominently the decoherence caused by environmental noise.

\section{Mechanisms for Creating and Distributing Entangled Qubtis}
\bigskip

\bgroup
\fixFloatSize{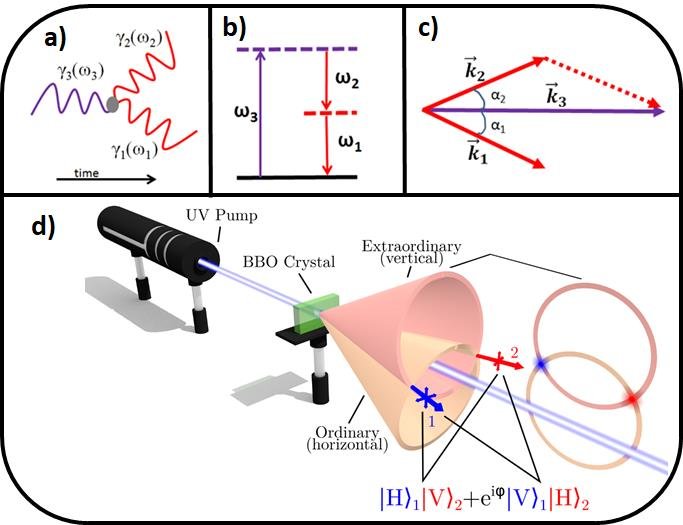}
\begin{figure}[!htbp]
\centering \makeatletter\IfFileExists{images/a8efacd2-cfc2-4e2a-b2ff-48ba54d534f8image23.jpeg}{\includegraphics{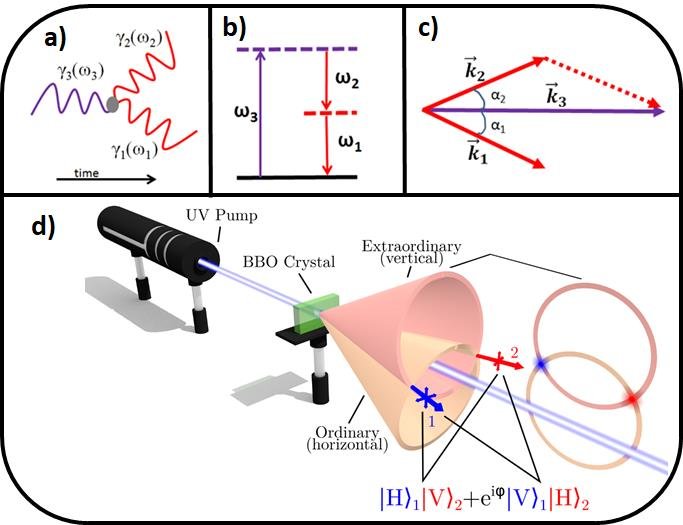}}{}
\makeatother 
\caption{{a) Feynman diagram of SPDC b) Energy conservation before and after emission c) momentum conservation before and after emission d) Schematic of SPDC as performed in the lab \unskip~\cite{couteau}.}}
\label{figure-10a38284ed4841b08b04743cd1db05d6}
\end{figure}
\egroup

Quantum communication is currently implemented using photons. These have the advantages of being easy to produce and to send to different locations, and that they can be manipulated with fairly high precision by optical components without separating from their entangled state. However, their lack of interaction makes photons difficult to entangle in arbitrary states. Currently, bipartite photon entanglement is easy to produce through a mechanism known as spontaneous parametric down conversion (SPDC), but creating entangled states of more than 2 photons is very difficult.

To produce entanglement by SPDC requires interaction of a primary photon stream at a given wavelength, $\lambda, $with a crystal with a second order nonlinearity; the primary photons are absorbed by the crystal and down-converted to a stream of two entangled photons at twice the wavelength, $2\lambda, $of the primary photons (Figure 9). The created photon pairs are called the signal and idler, and in most experimental set-ups are degenerate. Conservation of momentum requires that the emitted photons have opposite spin, creating entanglement. This can currently be done at high data rates of 1.5 Mb/s by creating an intermittent static electric field in a fiber and sending photons through it.

After distributing photons, it is important to correct their wavefronts so that they can be properly detected, but without introducing noise that can separate the photons. This requires adaptive optical components which do not interfere with the signal.

 Nuclear Magnetic Resonance (NMR) uses radio frequency resonance to control the spin of electron fields. These spins can act as qubits, and their interactions allow quantum communication\unskip~\cite{703868:16682649}. In particular, NMR has demonstrated much of what can be achieved without pure states or entanglement. However, it is important to note that NMR does not scale well, with additional qubits requiring an increasing amount of resources\unskip~\cite{703868:16682649}. Thus, while NMR will not itself be used for communication, it plays an important role in demonstrating the results of theoretical communication techniques.

\section{Entanglement-Based Protocols}
\bigskip

\textbf{QKD Protocols}

\subsection{Decoy State QKD}

Figure 8 shows the ability of an eavesdropper to carry out a "photon splitting attack" when an imperfect source sends more than one photon per pulse. This creates a breach in security, as the photons in a single pulse will have identical polarization. In response, the initial solution was to run sources at ultra-low power to limit production of photons to a low density. However, this resulted in an unacceptably low rate. In 2004, use of decoy states was proposed by H. K. Lo et al. as a solution to this problem\unskip~\cite{703868:16682655}. Decoy states are photons sent specifically for the purpose of testing transmittance. If an eavesdropper is intercepting photons, the receiving rate of decoy states will differ from what would be predicted based on the rate of signal states. Decoy states operate at different power levels from the signal states and therefore have a different photon number probability distribution. By comparing the loss of decoy states to the signal, loss of the different photon numbers can be estimated. This can reveal if an adversary is splitting off photons, thus improving secure state rates with simpler equipment.

Currently, all single source non-entangled protocols use decoy states to estimate loss. Since channel parameters like transmission loss and QBER must be estimated regardless, decoy states can perform multiple roles and thus improve transmission rates across the board.

Protocols based on decoy states are currently the best performing, at several Mb/s over distances of about 50 km of fiber. Even at higher noise levels, these protocols perform well with sending rates of several kb/s over distances of 150 km of fiber\unskip~\cite{703868:16682636}. However, at very short distances they are outperformed by entanglement-based schemes, which can leverage a greater number of bits per photon. Since decoy state protocols do not use entanglement, characterizing their noise is straightforward: this can be done by describing the QBER and transmission. Secure systems usually require a QBER of less than 10\%, which is less than what could be introduced by optimal eavesdropping.

Decoy state protocols are point-to-point communication schemes, which can introduce significant practical challenges. First, they require that two users share a quantum channel to send signals. If they are separated too far apart, they may need to rely on intermediaries (i.e., repeaters) to relay the key. However, this requires trust, which is undesirable in a cryptographic system because it opens up additional channels that could be attacked. In addition, in a network of n users, the number of links between users that is required to generate keys scales quadratically. This is far worse than protocols that scale linearly, such as entanglement-distribution schemes.

\subsection{High-Dimensional QKD}

Use of time-energy entangled photons leads to a large alphabet of time bins. This idea was originally proposed by Bechmann-Pasquinucci et al. \unskip~\cite{703868:16682623} and was recently realized by Author et al\unskip~\cite{703868:16682647}, achieving a secure key rate of Mb/s and 6.9 bits/photon on average.

This protocol is the first entangled photon protocol to be competitive with decoy state protocols, reflecting progress in entanglement generation, and the freedom that it allows in information encoding. However, the high rates were only achievable at extremely low noise at short distances. Entangled photon pairs are easily separated. In addition, if neither photon is detected, the bits are not sent, which means detector efficiency affects the rate quadratically.

The performance of this protocol in relation to its channel is characterized by timing jitter, which can cause a photon to end up in the wrong time bin. However, as shown by Zhong et al. \unskip~\cite{703868:16682654}, the timing can survive significant decoherence in a channel that would normally break entanglement. The main drawback of this protocol is that it requires one of the parties to send the entangled photons, and thus it does not apply to situations where entangled pairs are distributed by a third party.

\subsection{Continuous Variable QKD (CVQKD)}

As we will discuss in the following, some protocols go beyond the discrete |0\ensuremath{\rangle}/|1\ensuremath{\rangle} representation of quantum states by considering measurements that can take on continuous variable values. For example, the length of time taken before an entangled photon is detected can be entangled with the measured energy of the second photon. These time/energy measurements are continuous values with correlated Gaussian distributions. The correlation may be more sensitive to noise but it can generate far more than one bit of shared information.

Rather than measure the state in a discrete |0\ensuremath{\rangle}/|1\ensuremath{\rangle} basis, one might decide to measure properties which can take on a continuous range of values. For example, the time-to-detection and the energy of photons can be entangled. This would result in our official communicants, Alice and Bob, gaining correlated Gaussian variables. In a typical CVQKD protocol, when sending a continuous variable, Alice would prepare an EPR state and make a heterodyne measurement of the first qubit. She then sends the other qubit, embodied by a photon, to Bob, who makes a homodyne measurement. These different measurements correspond to energy/time measurements. After informing Alice of what observable he obtained, both Alice and Bob now have Gaussian variables correlated to one another, which can be used to generate several bits of the private key.\unskip~\cite{703868:16682633}

This particular protocol has achieved a secure transmission rate of 1 Mbit/s, which is below the current state of the art but still reflects significant progress. CVQKD needs a low noise environment but does not require use of sophisticated detectors. Homodyne detection is more consistent and cheaper to implement than single photon detection, suggesting it may find long-term use in short-range applications.

When the CVQKD protocol is in use, a channel through which information is transmitted is characterized by the excess noise, which is the variance introduced to the variable by the quantum channel. The excess channel noise can be used to bound the entropy retained by the state.

\subsection{Entanglement Distribution}

\bgroup
\fixFloatSize{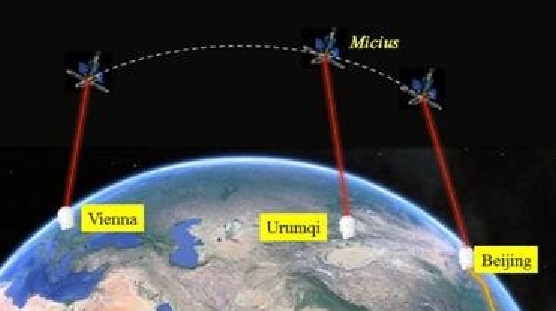}
\begin{figure}[!htbp]
\centering \makeatletter\IfFileExists{images/a8efacd2-cfc2-4e2a-b2ff-48ba54d534f8image24.jpeg}{\includegraphics{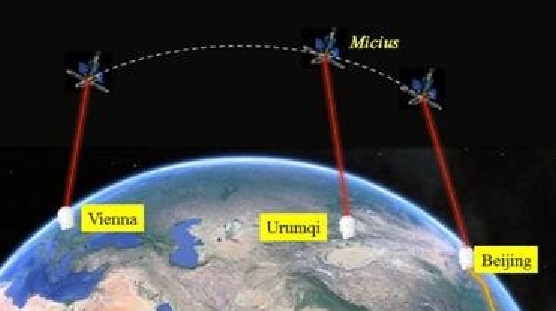}}{}
\makeatother 
\caption{{Distribution of entanglement between continents using satellites \unskip~\cite{sat}.}}
\label{figure-a746a35c0e9b434eb2da5088f6ff0aa5}
\end{figure}
\egroup

A central entangled source, accessible by many parties, could enable a large - scale secure quantum network. As described in previous sections, in a quantum communication system, Alice and Bob can verify entanglement of their states since monogamy of entanglement guarantees their states are maximally entangled. Without trusting a third party distributor, they can create a shared secret key, or perform whatever other form of quantum communication they want.

Such a network eliminates the need for a physical quantum channel between Alice and Bob, as teleportation can achieve whatever communication is needed. Furthermore, the number of actual physical channels would be one per user, from a user to the central distributor. This is in contrast to the quadratic scaling of simple networks and allows a more efficient management and compensation of channel parameters.

In 2017, researchers in China\unskip~\cite{703868:16682639} used a satellite to distribute entangled photon pairs across continents (as shown in Figure 10). This development has the potential to enable entanglement-based quantum communication channels worldwide, without the need of noisy physical infrastructure that requires transmission media such as fibers. This would allow the distribution of entangled pairs between distant locations, as was demonstrated between Beijing and Vienna. By removing the noise in physical media, this would allow quantum communication on a larger scale than would otherwise be possible.

\subsection{Measurement Device Independent QKD (MDI-QKD)}

\bgroup
\fixFloatSize{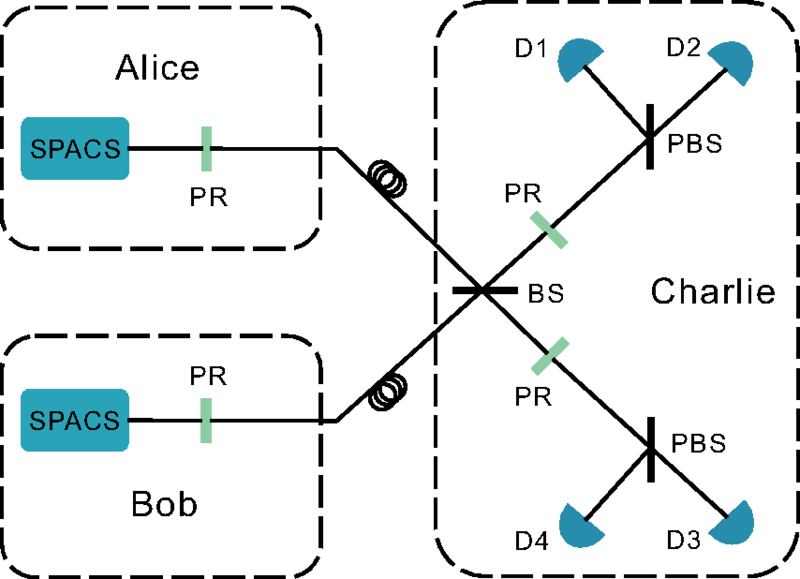}
\begin{figure}[!htbp]
\centering \makeatletter\IfFileExists{images/a8efacd2-cfc2-4e2a-b2ff-48ba54d534f8image25.png}{\includegraphics{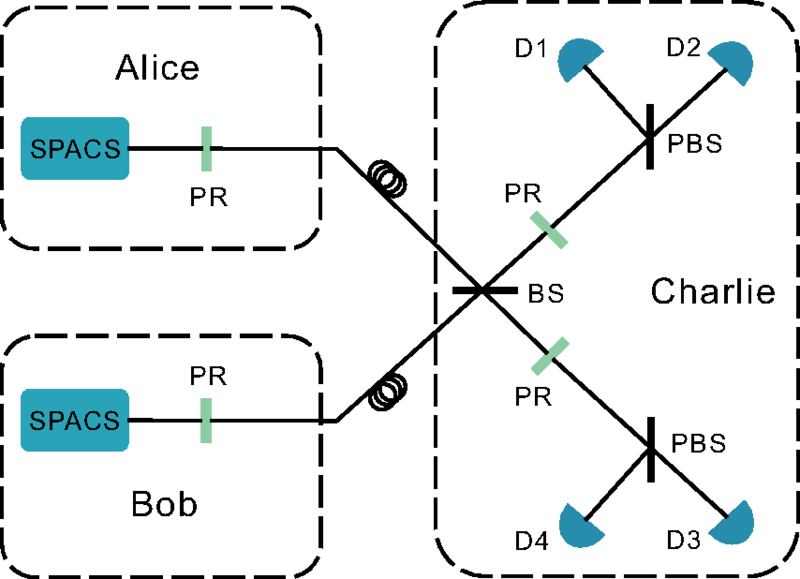}}{}
\makeatother 
\caption{{A diagram of the MDI-QKD scheme. Alice and Bob, using Single Photon Added Coherent Sources (SPACS) choose their polarization basis (using the polarization rotator PR) randomly. Charlie measures the photons by first having them interfere at the beam splitter (BS), projecting them onto the horizontal/vertical basis with the polarizing beam splitter (PBS), and then reading the response of the detectors (D1 through D4). Charlie communicates which two detectors detected a photon. Based on the results Charlie announces, Alice and Bob can determine a shared key with their secret knowledge of the bases \unskip~\protect\cite{703868:16682646}.}}
\label{figure-5b5bafd0ce4041a7b124c87f28622ce5}
\end{figure}
\egroup

In yet another type of protocol that is different from device-independent quantum key distribution, measurement device independent (MDI) quantum key distribution allows Alice and Bob to send photons using random bases to a central party that performs measurements. Bob then announces the measurements which permit both Alice and Bob to share correlated information on their basis choices without disclosing the same information to the central party that does the measurements (see Figure 11). Such a protocol avoids the difficulty of having expensive single photon measurement devices that must be kept at extreme cryogenic temperatures.

The MDI-QKD protocol has allowed transmission of discrete (|0\ensuremath{\rangle}/|1\ensuremath{\rangle}) variables across very long distances, and is a strong candidate for satellite supported QKD\unskip~\cite{703868:16682652}. However, the transmission rates for this protocol are rather low. Nevertheless, when this scheme is utilized with continuous variables, it is extremely fast and has low noise, for example in short-distance optical networks. There is therefore a potential to use this protocol for geographically close nodes connected by fiber.

Table III gives a concise summary of the QKD schemes discussed. The main takeaway is that the Decoy State QKD methods provide the best results in terms of cost, robust setups, and maximum bit rate distance products. In the future, more noise-sensitive methods like high-dimensional continuous variable protocols might yield the best maximum bit rate distance products.

\textbf{Channel Usage}

Correlations between entangled states can allow improvements on shared channel usage. A typical channel usage protocol is the Slotted Aloha system. This allows users to obtain up to 37\% channel usage (or throughput) without central coordination\unskip~\cite{703868:16682656}. However, if users share an entangled state, they can perform measurements to receive a correlated result. For example, two users sharing an EPR pair could measure in the |0\ensuremath{\rangle}/|1\ensuremath{\rangle} basis, agreeing beforehand that whoever measured |0\ensuremath{\rangle} would communicate at ontime, and |1\ensuremath{\rangle} at another. Thus, without ever communicating with each other, they know what time they should use the public channel. This can allow up to 100\% channel usage, which is nearly three-fold of what is considered optimal in classical slotted-ALOHA protocol. Interestingly, this allows to approach a 100\% throughput performance, similar to the well-known and widely used Carrier-Sense Multiple Access (CSMA) schemes.

\textbf{Quantum Communications and} \textbf{ Distributed Quantum Computation}

Using quantum teleportation, users performing quantum computations at different locations could share important qubits without observing them. This requirement means that a global quantum Internet such as that proposed by Liao\unskip~\cite{703868:16682659} could enable global cooperation on quantum computation.

\begin{table*}[!htbp]
\caption{{Overview of Pros and Cons of QKD schemes and Best Results Reported So Far \unskip\cite{703868:16682652}.}}
\label{table-wrap-c796c8635dd54f67913544a36742821e}
\def\arraystretch{1}
\ignorespaces 
\centering 
\begin{tabulary}{\linewidth}{LLLL}
\hline 
 PROTOCOL &
   PROS &
   CONS &
   MAXIMUM BIT RATE DISTANCE PRODUCT ACHIEVED (Mb/s * km)\\
   \hline
 Decoy State QKD &
      \ensuremath{\bullet{}} Easy state preparation
      \newline\ensuremath{\bullet{}} High noise tolerance
      \newline\ensuremath{\bullet{}} High rate with cheap parts
      &
      \ensuremath{\bullet{}} Requires quantum channel
      \newline\ensuremath{\bullet{}} Quadratic network scaling
      \newline\ensuremath{\bullet{}} Limited maximum rate and distance
    &
   2.5\\
   \hline
 High-Dimension QKD &
      \ensuremath{\bullet{}} High information per photon
      \newline\ensuremath{\bullet{}} Highest rate at low noise
      \newline\ensuremath{\bullet{}} Correlation survives decoherence 
      &
      \ensuremath{\bullet{}} Limited by square of detector efficiency
      \newline\ensuremath{\bullet{}} Poor at high noise
      \newline\ensuremath{\bullet{}} Requires multiple detectors
     &
   2.7\\
   \hline
 CVQKD &
      \ensuremath{\bullet{}} Continuous variables allow for a larger alphabet
      \newline\ensuremath{\bullet{}} More bits per photon
      \newline\ensuremath{\bullet{}} Simple detectors
    &
      \ensuremath{\bullet{}} Poor at high noise
      \newline\ensuremath{\bullet{}} Low rates achieved
    &
   1\\
   \hline
 Entanglement Distribution &
      \ensuremath{\bullet{}} Simple scaling to new users
      \newline\ensuremath{\bullet{}} Enables protocols beyond QKD
      \newline\ensuremath{\bullet{}} Flexibility in protocol
    &
      \ensuremath{\bullet{}} Requires central distributor 
      \newline\ensuremath{\bullet{}} Poor at high noise
    &
   N/A\\
   \hline
 MDI QKD &
      \ensuremath{\bullet{}} Doesn't require Alice and Bob to have measurement devices
      \newline\ensuremath{\bullet{}} Possible at high noise
      \newline\ensuremath{\bullet{}} Can use continuous variables
    &
      \ensuremath{\bullet{}} Low rate at high noise
      \newline\ensuremath{\bullet{}} Requires central party for measurement
      \newline\ensuremath{\bullet{}} Single use
    &
   2\\
   \hline
 DI QKD &
      \ensuremath{\bullet{}} Doesn't require users to have photon sources
      \newline\ensuremath{\bullet{}} blah
    &
      \ensuremath{\bullet{}} Quadratic dependence on detector efficiency
      \newline\ensuremath{\bullet{}} Low data rate at high noise
      \newline\ensuremath{\bullet{}} Requires central party for signal
    &
   N/A\\
\hline 
\end{tabulary}\par 
\end{table*}

\section{Beyond QKD}
\bigskip

\subsection{Quantum Relays}

It should now be clear that many protocols that use quantum entanglement are limited to very short distances because of noise. In classical communications, repeaters provide a simple solution to this problem. As a signal deteriorates over distance, it can be regenerated and re-transmitted at higher power\unskip~\cite{703868:16682644}. In this way, errors due to noise corruption can be kept low so that the majority of bits are preserved. In general, a large number of photons can be used for signaling as there is no assumption that the message has not been eavesdropped on.

Quantum communication, however, cannot solve this problem in the same way as is done in classical communication. Quantum signals are usually very low power, comprising few photons. Furthermore, signals cannot be regenerated and re transmitted because the quantum essence of information is lost. Measurement can only be performed at the receiving end.

A proposed possible solution is the quantum relay\unskip~\cite{703868:16682631}, comprised of a set of nodes that can generate entanglement between two other nodes. Such as scheme would allow quantum signals to be teleported from one node to the next, effectively shortening the amount of distance actually travelled by the signal. This experiment was performed in 2005 and a 77\% fidelity over 2 km of fiber\unskip~\cite{703868:16682653} was achieved, which is better than what could be obtained without entanglement.

Quantum relays such as the one described above have been shown to be secure in QKD schemes; they could, in principle, extend achievable distances to hundreds of kilometers if modern detectors are used\unskip~\cite{703868:16682637}. Using quantum relays has also been shown to be effective in avoiding detector side-channel attacks compared to other schemes\unskip~\cite{703868:16682650}. They have been implemenented on-chip, which in future systems could allow the higher fidelity communications required for quantum computations \unskip~\cite{703868:16682641}.

\subsection{Quantum Repeaters}

Quantum repeaters are essentially elaborate quantum relays, which allow entanglement distribution across large distances. By generating entangled pairs at every link on a chain of stations, teleportation can be used to transfer entanglement across greater distances than relays are capable of\unskip~\cite{703868:16682658,703868:16682627}. However, quantum repeaters and the process involved require quantum memories to store the quantum state until it is needed. The coherence of this memory is critical to the success of repeaters. There has been limited experimental success on this front\unskip~\cite{703868:16682651}.

\subsection{Quantum Memory}

Even though quantum memory is important for synchronization and gate implementation in quantum computers\unskip~\cite{703868:16682629}, it is also relevant to quantum communications. This is currently an active area of research and experimentation. State-of-the-art experiments have achieved limited success, with 90\% fidelity at low transmission rates\unskip~\cite{703868:16682626}. Given the slow progress, it is unlikely that quantum repeaters will achieve the level of success currently enjoyed by satellites in quantum communications within the next 10 years.

\subsection{Quantum Multicasting}

Multicasting is the process of sending a signal to multiple receivers. It is desirable in a cryptographic network to provide secure classical multicast. However, little work has been done in this area. It has been shown that use of multicast quantum states lowers quantum channel capacity\unskip~\cite{703868:16682648}. The typical network architecture used for quantum multicast is not the same as that in classical network multicast scenario since single qubits may not be replica ted when sent to different parties\unskip~\cite{703868:16682632}.

\subsection{Quantum Broadcast}

Broadcasting is the process of sending a message to an entire network. In quantum communication, this would allow communication between any pair of parties in the network, or secure communication to all parties. This has been proposed by Yard using GHZ states\unskip~\cite{703868:16682661}. However, due to measurement effects, quantum broadcasting may require sending many signals, or, in the worst case, repeating the message to each receiver in a point-to-point manner. By extension of the no-cloning theorem, a quantum state cannot be broadcast in two separate systems, even if only marginal reproduction is needed\unskip~\cite{703868:16682635}. This limits what kind of protocols could be achieved.

One use of broadcasting protocols is to provide optimal bounds for more general quantum communication. There is interest in using broadcasting in trusted networks, or where eavesdropping is not a concern\unskip~\cite{703868:16682645}.

\section{Observations and Recommendations}
\bigskip

Given the cost and challenges of creating, maintaining, and measuring entanglement, the key question that is frequently asked is: is entanglement-based communications worth the cost? For some areas, it remains to be seen if costs and limiting technologies will reach the point of practical usefulness. However, several immediate and near-future applications can only be fulfilled by entanglement-based communications, meaning there will be a significant need for additional research and improved architectures for quantum communications.

\subsection{Resilience to Quantum Computing}

The fundamental motivation for this paper is the high probability that quantum computing will be able to break current cryptographic methods in the next decade. It is conceivable that Quantum Computing will be used by several governments and big companies in the near future. Quantum computing will be capable of efficiently disrupting the secrecy of classical cryptographic methods, while quantum cryptography will never be broken given the current laws of physics. Any current environments which require long-term security will need to use quantum cryptography in preparation. For example, the interesting experiment performed in Switzerland using QKD and Quantum Cryptography was carried out in cooperation with a bank as banks currently carry out a large volume of sensitive communication \unskip\cite{nilsson}. In addition, it would be prudent to prepare for the increase in quantum computing power by having a cryptographic system in place which is agnostic and resilient to the advances in quantum computing and cannot be broken.

\subsection{Unbreakable Security}

Even if it takes more time for quantum computing to become mainstream, it appears that one of the immediate applications and potential markets for quantum communication might be for environments requiring long-term unbreakable security. For example, classified documents in the US may remain so for up to 50 years, a period too long to predict technological or mathematical discoveries. Currently, these top-secret documents have strict limitations on being sent electronically. This could be bypassed by a QKD system which is provably secure, and less susceptible to side-channel attacks. As demonstrated by the Micius satellite, this could be accomplished by a satellite network with entanglement-distributing capabilities. This would even allow classified documents to be sent overseas.

\subsection{General Quantum Communications}

An effective system for distributing entangled states will also allow various types of non-encrypted quantum communications. For example, this will allow labs in different locations to teleport quantum states to each other. This, in turn, would allow labs with different equipment to carry out different parts of an experiment, broadening the range of viable quantum experiments which could be carried out cooperatively. As another example, such distribution would allow the testing of superdense coding techniques to allow an exponential amount of bits to be sent with the limited number of photons used in quantum communications.

\subsection{Adaptive Topology}

Quantum cryptography will also be an effective solution in networks which require adaptive topology. This is another significant dimension of network security and secure communications. In other words, there are several instances where malicious security attacks might necessitate an adaptive topology to ensure continuity of secure network communications between different nodes of a network since one or more of the original nodes might be compromised. 

For example, a simple star network in which a central hub can communicate with all other nodes prevents communication between nodes without passing through the hub. However, if the hub distributes entangled pairs between the nodes, they will then be capable of forming a virtual link using quantum teleportation. This could be the case in a satellite enabled network in which ground stations receive entangled states from a single satellite. Hence, it is clear that quantum entanglement could provide an effective defense mechanism against malicious security attacks by supporting adaptive topology in a given network. This means that the topology of a given network can be modified on the fly to bypass the compromised nodes or portions of a network. 

\subsection{Bandwidth Storage}
As discussed in Section III, superdense coding would allow one to convert distributed entangled states into classical communication. A system of n qubits would allow a number of bits exponential in n to be sent. This method could be used to "store" bandwidth in limited environments with high peak demand \unskip\cite{giovannetti}. By distributing entangled systems during a period of low demand, a large number of bits can be sent in a short period of time, beyond the normal capacity of a channel. Such a technique would require that parties be able to perform the required measurements, and to store the entangled states for sufficiently long. While this application would be of great use, current entanglement distribution methods are too lossy to yield any gains. If low noise methods of entanglement distribution and storage were developed, this technique could see widespread adoption, as a common challenge for modern communication networks is peak load \unskip\cite{chlamtac}.

Thus, if unbreakable security is required, quantum cryptography and quantum entanglement is the ultimate frontier as it is expected to achieve guaranteed security in encryption at the highest rates. In particular, entanglement works best in low noise environments where high-dimensional encoding can be used. In noisy environments, or where security requirements are more lenient, the future use of quantum cryptography remains to be seen.

For the Internet in general, only the pseudo-secure rate matters. It is not clear that this rate will be improved by quantum communication any time soon. Quantum computing is not yet able to threaten communications, so if long-term security isn't a strict requirement there might not be reason for concern.

The future use of quantum communication through noisy environments is also uncertain. Long-distance free space communications have shown more limited results than through fiber, due to atmospheric noise. It seems that a more robust solution might be to achieve free space communication through a satellite, which minimizes the distance travelled through the atmosphere.

It appears that, in principle, several potential applications require the use of quantum relays or long-term quantum memories. These technologies have attracted significant interest and research but it remains to be seen whether a practical version will arise in the near future. For example, exploiting superdense coding will require parties to share many entangled states beforehand. Until these states can be reliably preserved, such a scheme might not be practical.

The ability to broadcast cryptographic keys will also be a challenge. Due to the non-cloning theorem, many quantum states may have to be prepared to send a message to as many parties. The creation and stability of high dimensional quantum states is an ongoing challenge and the viability of "true broadcast" on a quantum network will depend on progress in manipulating such states. An interesting area of research will be how best to carry out broadcast when limited by current quantum technologies.

\section{Related Work}
\bigskip

In this section, we briefly review other survey papers and related work.

Frank Wilczek's paper provides a simple and excellent overview of what distinguishes classical systems from entangled systems\unskip~\cite{Wilczek}. It also covers different examples without classical descriptions and what assumptions must be made for quantum vs. classical explanations.

Gyongyosi et al. surveys the fundamental differences between classical and quantum channels\unskip~\cite{703868:16682654}. The key concepts and metrics of quantum channels are laid out with significant mathematical detail. Quantum channels admit many possible definitions of capacity and have a richer array of properties. This paper discusses the capacities and properties of quantum channels, and the differences between classical and quantum channels.

Enriquez et al. surveys the notions of maximal entanglement for higher dimensional composite quantum systems. The paper emphasizes the importance of which measure is chosen\unskip~\cite{703868:16682643}. For example, the GHZ and W states are both maximal under different metrics. The paper addresses identifying an entangled state and presents numerical results on the maximal entanglement of 4-party states.

Broadbent et al. covers cryptographic methods using quantum mechanics other than QKD, such as quantum money, random number generation, multi-party computation and quantum computation delegation, bit commitment, authentication, and quantum rewinding\unskip~\cite{703868:16682642}. Some of these techniques are enabled or enhanced by entanglement.

Diamanti et al. covers Continuous Variable QKD (CVQKD) (i.e., coherent implementations) schemes\unskip~\cite{703868:16682652}. The paper goes over the different protocols, experimental implementations, security proofs and concepts involved in using coherent states for QKD. While the paper is not comprehensive, it aims to provide a better understanding CVQKD. It covers both DI and MDI QKD.

Dimanti et al.\unskip~\cite{703868:16682638} covers the limiting factors to different QKD metrics like cost, distance, size, and key rate, as well as practical security. It discusses hardware limitations, covering on-chip systems, current emitters and detectors, and channels used. It also discusses software limitations of protocols, both point-to-point and DI/MDI.

All{\'e}aume covers the different applications of QKD for enabling more secure symmetric encryption, such as AES\unskip~\cite{703868:16682640}. These techniques aren't information-theoretically secure; such a scheme replaces public key methods like RSA with QKD. The paper also covers point-to-point vs. network QKD protocols, describing trusted networks and device-independent protocols.

Gisin et al. gives an introduction to QKD and covers protocols from 1 to 4 photons; i.e., point-to-point, entangled, teleportation, and entanglement swapping protocols\unskip~\cite{703868:16682625}. The review covers idealized forms of quantum communications, without addressing channel capacities or going far beyond QKD.

\section{Conclusions}
\bigskip

Quantum entanglement offers a new modality for communications that is different from classical communications. The motivation for use of quantum entanglement is its flexibility and security even though some of the quantum entanglement protocols are non-intuitive. It seems clear that traditional communications engineers will need to develop a new way of thinking that moves them away from classical intuition in order to take advantage of the entanglement resource in building advanced communication systems.

Using entanglement in communication systems requires a choice on what kind of state to prepare, and what measurements to perform. These choices should be made not only on the basis of what is possible, but also on what sort of correlation is desired. Generally, measurements performed determine what sort of information is shared, which form the main attributes of any protocol. Unlike in classical communications, where understanding what information is transmitted or manipulated is straightforward, this is not so in quantum communications where quantum entanglement might play an important role. In quantum communications, the measurements required determine what sort of states one can distribute, and what metrics are used to quantify the channel noise that is most relevant. Noise generally results in the decoherence of the states, but the effect depends on what type of measurement is performed.

It is important to remember that entanglement can be verified and distilled to a pure form. This means that, given the right infrastructure for distributing photon pairs, parties can perform the communications they want even at modest rates. There are no theoretical barriers to enabling a global quantum network, where parties without quantum channels are able to communicate, as entangled pairs will allow parties with a classical channel to perform quantum communication via entanglement swapping.

The performance of entanglement-based QKD schemes have recently begun to outperform the older and simpler BB84-type protocols, and unlike the single-use BB84, entanglement-based communication has many uses. The flexibility of quantum entanglement means that a global network could see many applications beyond cryptography. As quantum computing begins to threaten the security of traditional cryptography, we are likely to see investment in such a network in the near future.

\section*{Acknowledgement}
\bigskip
This work was supported in part by a Dean's Fellowship for Keith Shannon.



%

\bibliographystyle{IEEEtran}

\bibliography{article.bib}

\end{document}